# Using Cognitive Agent-based Simulation for the Evaluation of Indoor Wayfinding Systems

*A Case Study*

H. Schrom-Feiertag and M. Stubenschrott and G. Regal and J. Schrammel and V. Settgast[1]

*AIT Austrian Institute of Technology*
*[1]Fraunhofer Austria / Graz University of Technology*
*Giefinggasse 2*
*1210 Vienna*
*Austria*
*helmut.schrom-feiertag@ait.ac.at*

**Abstract**: This paper presents a novel approach to simulate human wayfinding behaviour incorporating visual cognition into a software agent for a computer aided evaluation of wayfinding systems in large infrastructures. The proposed approach follows the Sense-Plan-Act paradigm comprised of a model for visual attention, navigation behaviour and pedestrian movement. Stochastic features of perception are incorporated to enhance generality and diversity of the developed wayfinding simulation to reflect a variety of behaviours. The validity of the proposed approach was evaluated based on empirical data collected through wayfinding experiments with 20 participants in an immersive virtual reality environment using a life-sized 3D replica of Vienna's new central railway station. The results show that the developed cognitive agent-based simulation provides a further contribution to the simulation of human wayfinding and subsequently a further step to an effective evaluation tool for the planning of wayfinding and signage.

# 1.  INTRODUCTION

The architecture of public buildings like train stations and the provision of guidance information has important implications for people, regarding the quality of the provided services and constitute a major challenge for designing a consistent and accessible wayfinding and signage system. The locations for visual signs need to be selected carefully such that they do not compete with other signs or advertising, are readily viewable within an adequate time, are not partially obstructed, and can be seen within viewing distances of an average person. All these aspects of a wayfinding and signage system should be taken into account already in the planning phase. Wayfinding has to be treated as an integral part of the design process to create more intuitive architectural spaces where people navigate instinctively. Currently, there are no adequate tools available that allow an integration and assessment of such issues into the planning process.

Therefore we propose a new agent-based simulation model enabling a computer aided evaluation of wayfinding systems to identify weaknesses and gaps in the signage system of an infrastructure. In this work, we focus on the simulation of wayfinding of people who are unfamiliar with an infrastructure and are heavily dependent on available signage. As test scenario we used Vienna's new central railway station, as this provides a prototypical example for complex wayfinding problems with a focus on pedestrians. Train stations gradually evolve from transport hubs towards multi-functional environments combining mobility services with commercial areas for events, exhibitions, shopping or various other activities. This increases the complexity of the environment which can cause disorientation and discomfort.

The paper is structured as follows. In Section 2 the state of the art in pedestrian simulation, wayfinding and visual attention modelling is presented. Section 3 describes the developed wayfinding simulation with the different models and how they are cooperating. In Section 4 we describe wayfinding experiments where we gathered empirical data by means of a virtual reality (VR) environment. Then we compare the results from the agent-based simulation with the empirical data. The paper closes with concluding remarks and discussion on future challenges in Section 5.

# 2.  RELATED WORK

## 2.1  Agent-based Pedestrian Simulation

Simulating pedestrian behaviour has recently gained a lot of attention in a variety of disciplines, including urban planning, transportation, civil engineering and computer science. Various kinds of models like cellular automata, fluid dynamics, discrete choice models, rule-based models and multi-agent models for simulating pedestrian behaviour have been suggested (Timmermans, 2009). Agent based microscopic modelling is an approach for simulating pedestrians as single individuals by supplying a detailed representation of their behaviour, including decisions on various levels and interactions with other pedestrians in the crowd with the goal to reproduce realistic autonomous behaviour. Pedestrian behaviours can be categorized as strategic, tactical and operational behaviour as referred to in (Bierlaire and Robin, 2010, Hoogendoorn, Bovy, et al., 2002, Hoogendoorn and Bovy, 2004, Kielar and Borrmann, 2016). Strategic behaviour describes destination choice and activity scheduling, tactical models characterize the pedestrians' route selection from the pedestrian's current position to a certain destination under which wayfinding can be understood. The operational behaviour relates to the manner of walking to the next visible intermediate goal of the route and interacting with other pedestrians and obstacles along the path.

Most currently available simulation models are based on the assumption that all pedestrians know the infrastructure perfectly and consequently all pedestrians choose the shortest path to reach their goal. However, for those pedestrians who are not familiar with the infrastructure a more realistic simulation of the wayfinding behaviour is needed.

## 2.2 Wayfinding Behaviour

In (Koh and Zhou, 2011) the important factors in a pedestrian's decision-making process during wayfinding include a pedestrian's sensory attention, memory, and navigational behaviours. Wayfinding through buildings like train stations can be considered as route following facilitated by signs, easily associable landmarks like shops and crossings connected by corridors and can be classified after the taxonomy in (Wiener, Büchner, et al., 2009) as aided wayfinding. Aided wayfinding is considered to be rather simple since it does not require considerable cognitive effort from the user. It is important only to provide all the relevant information at each decision point. Research on people's wayfinding helped to establish practical guidelines on how to design public buildings and signage to facilitate wayfinding (Arthur and Passini, 1990). Large public buildings such as central railway stations fulfil various functions, which makes the design of signage and wayfinding systems very difficult and error-prone. Currently, there are no adequate tools available that allow the assessment of signage systems already in the planning stage.

The approach of applying computational cognitive models for understanding human cognition is relatively new and significant progress has been made in recent decades in advancing research on computational cognitive modelling. But, there is still a long way to go before we fully understand the computational processes of the human mind (Sun, 2008). There exist various computational cognitive models for wayfinding which are focussing primarily on the exploration of mental representations rather than on the information needed for wayfinding and neglect the processes of how people perceive and navigate through spatial environments (Raubal, 2001).

Therefore it will be necessary to carry out further studies to advance the understanding of human wayfinding behaviour. Recent developments in virtual reality head mounted displays and eye-tracking enable new innovative possibilities and will boost wayfinding research which will provide in-depth insights to human behaviour.

## 2.3    Visual Attention

Understanding the visual attention process of pedestrians during wayfinding tasks is an important prerequisite to successfully and realistically model their behaviour. Different important aspects of the process must be addressed: Visual attention needs to be understood and analysed as a result of both bottom-up and top-down processes.

Each visual scene has its own visual properties, and some visual patterns attract more attention than others. (Itti, Koch, et al. 1998) analysed different images and showed that visual saliency can be used to model human attention towards images. This approach provides a helpful bottom-up characterisation of the visual scene, but research has shown that it can only account for a low percentage of fixations (Rothkopf, Ballard, et al., 2007).

In order to fully address visual attention processes also knowledge-driven gaze control needs to be considered. (Henderson and Ferreira, 2004) provide a typology of the different levels of knowledge involved. Episodic scene knowledge deals with information about a specific scene that is learned over the short term (current perceptual encounter) and over the longer term (across multiple encounters). Scene-schema knowledge relates to information about the objects likely to be found in a specific type of scene (e.g. train stations typically contain guidance signs), and spatial regularities (e.g. guidance signs are frequently placed overheads). Task-related knowledge describes a general gaze-control strategy relevant to a given task. For the example of navigating in a busy train station such a strategy might be to periodically switch between fixating the immediate walking path for collision avoidance and scanning overheads for guidance signs.

Recently, empirical studies relating attention patterns with body movements, head motion and gaze direction in naturalistic settings have been conducted (Foulsham, Walker, et al. 2011). More specifically data addressing the problem of walking were the people need to navigate through the environment and avoid collisions has been conducted by (Jovancevic-Misic and Hayhoe, 2009). Furthermore, more and more detailed attention data from realistic scenarios outside the lab become available through the use of advanced scene reconstruction and eye tracking equipment (Paletta, Santner, et al. 2013, Schrammel, Mattheiss, et al. 2011). These data now can be used to improve the quality of attention models for pedestrians.

## 3. AGENT-BASED SIMULATION

### 3.1 Overview

The main contribution of this work is a cognitive agent-based simulation which is tailored to the needs of persons unfamiliar with the infrastructure. To simulate human wayfinding in a plausible way, visual perception and cognition of guidance information need to be integrated. Therefore, the proposed wayfinding includes visual access to signage for directional information at decision points. It follows the Sense-Plan-Act paradigm (Gat, 1998) comprised of models for visual cognition of signage, navigation behaviour and pedestrian movement as explained in *Table 1* and described in the following sections.

*Table 1*. Basic algorithmic steps of the simulation loop.

```
Main Loop
    while target location is not reached:   # or the simulation time has run out
        perceive surrounding;                # render agent view in the virtual reality environment
        calculate attention                  # calculate probabilities of the visual attention for each sign
        select navigation behaviour;         # select appropriate navigation behaviour and action
        execute the movement;                # check collision and update agent's location
    end while
```

### 3.2 Modelling of the Environment

The virtual environment of Vienna's new central railway station was used for the visual view of the simulated agent as shown in *Figure 1a*. To determine the areas covered from signage in the field of view additionally a mask is rendered containing the signage only (*Figure 1b*). For each pixel in this mask a unique identifier (ID) to the associated sign is assigned (*Figure 1*c). So all signs and covered areas that are visible to the agent can be determined an evaluated. Semantic information about type and information provided by each sign can be retrieved from a database using the unique ID. Therefore a 3D editor was developed which made it possible to explore the 3D model and to select and annotate the signs with semantic information.

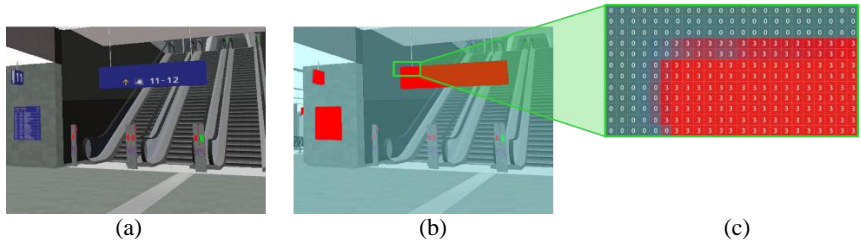

(a)                                      (b)                                       (c)

*Figure 1.* Rendered agent view (a), areas containing signs (b) and per pixel sign id's (c).

## 3.3    Modelling Visual Attention

The field of view rendered from the 3D model serves as input to the attention model. On the basis of this image the attention model calculates the saliency in the visual range using a frustum model (Riche, Mancas, et al., 2013). The result is an attention distribution on the objects in the field of view (Schrammel, Regal, et al., 2014).

As illustrated in *Figure 2*, a function for identifying objects of interest was developed. The function consists of three modules to calculate (1) the dynamic field of view (Frustum, Fru), (2) bottom up saliency (Sal) and (3) task specific attention (Semantic, Sem). The function for calculating objects of interest receives an image of the agents view and a mask that assigns an ID to each object for identification as input parameters. For simulating the frustum, a Gaussian distribution ($\mu=0$, $\sigma=7$) is combined with a Beta distribution ($\alpha=3$, $\beta=12$), based on the data by (Foulsham, Walker, et al., 2011). The visual saliency is calculated by using the RARE 2012 algorithm by (Riche, Mancas, et al., 2013). For the task specific attention a semantic model of relevance of objects is used with three types of objects relevant for navigation: signage, train schedule information, and other objects belonging to the train infrastructure (i.e. ticket machine). Related to the task the estimated attention for each object is calculated, based on attention defined in the model. The output of the three modules is combined by using a weighted geometric mean. This allows to simulate different user characteristic by applying a weight factor to the output of each model. Weighting factors were set as wSa=wSe=wFr=1.

The integrated task based attention is calculated as
$$\sqrt[wSa+wSe+wFr]{Sal^{wSa} + Sem^{wSe} + Fru^{wFr}}.$$

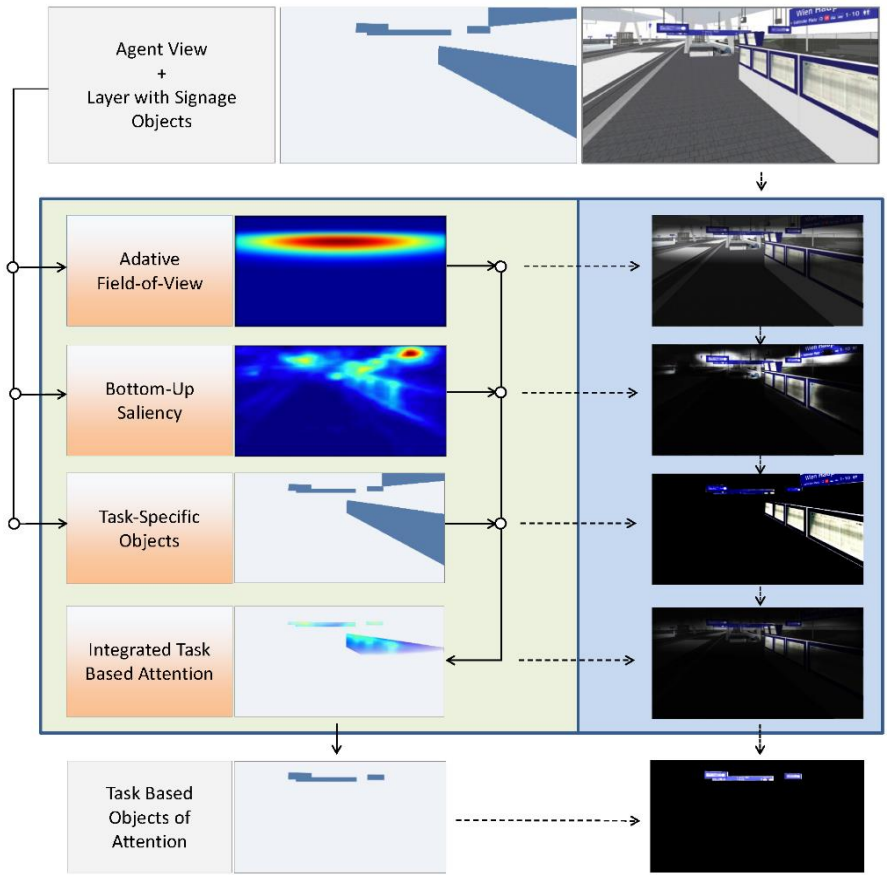

*Figure 2.* Task based visual attention model

Using the integrated task based attention, the most relevant objects are identified by calculating the attention for each object in the image separately. Therefore, the sum of attention of each point within an object is calculated and the objects are sorted by overall attention. The task based attention objects are provided as a sorted list to the agent navigation behaviour model.

## 3.4 Modelling Navigation Behaviour

The agent's behaviour is modelled based on the probability that each sign in the current field of view is seen and recognized, as calculated from the visual attention model. To enhance generality and diversity of the developed wayfinding simulation, stochastic features of perception are incorporated to reflect a variety of behaviours. Therefore, a fundamental aspect of the behaviour model is a *threshold(sign, agent)* function which defines the required attention threshold which must be exceeded in order to recognise the given sign. The value of this function is assigned randomly between 0 and 1 at the start of each simulation run for each agent/sign combination and remains constant during the simulation. For example, the same sign could have a threshold value of 0.3 for one agent and 0.8 for another. Signs with a good contrast in the centre of the agent's field of view have an attention close to 1 and the stochastic influence of this threshold can be neglected, whereas signs with a medium attention will only be seen by some agents.

When searching for a given target, the attention model calculates the attention for each sign in the field of view in regular intervals. If *attention(sign, agent) ≥ threshold(sign, agent),* the sign is added to the temporary set *S* of seen signs. The content of all seen signs is then matched with the given task and is classified into one of three categories: (1) The sign is directly at (or very near) the given target and it can be assumed that the agent knows the location of the target if he sees the sign. (2) The sign shows a clue in which direction to continue searching for the target (usually a direction arrow). (3) No suitable information can be found on the sign.

The signs are then sorted in terms of priority. Signs in category 1 are always preferred to signs in category 2, which in turn are preferred over signs in category 3. If there are multiple signs visible in the same category, the ones with a higher calculated attention are preferred. The final behavioural action is then inferred from the category of the sign with the highest priority. In case of (1), the agent is directly sent to the target and in case of (2), the agent is sent in the direction which is given by the sign. If only signs of category 3 or no signs at all could be seen, the agent continues searching for further information. This is currently done by sending the agent to certain base points which are manually placed in the infrastructure (green lines in *Figure 3)*. An overview of the general algorithm is given in the following *Table 2*.

*Table 2*. Decision algorithm of the behaviour model

```
bestSign = Ø                              # No sign selected as the best initially
S = get_signs_in_field_of_view(agent)     # Get a set of signs which are in the field of view
for s ∈ S:                                # Each sign in field of view
   if attention(s, agent) < threshold(s, agent):
      continue                            # Sign does not grab attention

   if bestSign == Ø or s.category > bestSign.category or
       (s.category == bestSign.category and attention(s, agent) > attention(bestSign, agent)):
      bestSign = s                        # Select the sign, if it has higher category than the current,
                                          # or as a fallback a higher attention
if bestSign != Ø:
   goto_position(bestSign.goal)           # Go to the given goal (or direction) from selected sign
else:
   explore_area()
```

### 3.5	Modelling Movement

The agent proceeds to the next point of interest obtained from the behaviour model. Points of interest can either be the target itself when it was found, or e.g. a point at the end of a hallway if a direction arrow on a sign points along that hallway. If no target (or clue to the target) was found, points of interest are scattered around the infrastructure and selected sequentially to explore the area. In all cases, the goal is a specific position in the infrastructure which the agent needs to reach with a movement model to navigate among neighbouring agents and obstacles through the VR environment.

Pedestrian motion is handled on two different levels. First, a path to the next point of interest is found by building a regular grid of the infrastructure and searching for the quickest path using the Theta* algorithm (Daniel, Nash, et al., 2010) which yields obstacle free waypoints to get to the goal. Second, the movement to each waypoint is modelled by a simple social force model after (Helbing and Molnar, 1995) where opposing forces from other pedestrians and walls are combined with an attractive force that steers the agent towards the next waypoint.

## 4.	EXPERIMENTS AND RESULTS

To evaluate the simulation results empirical data were collected from 20 participants by means of an immersive virtual reality environment using a life-sized 3D replica of Vienna's new central railway station. The main hall of the railway station is approximately 150 m x 350 m wide und comprises of three levels, 22 escalators and six elevators. The signage consisted of about 290 signs and details of placements and graphics of the sign were provided by the architect. We defined typical use cases to test our cognitive agent-based simulation for the evaluation of indoor wayfinding systems.

### 4.1	Scenario Description

The scenario covered a wide area of the train station and took 15 to 20 minutes for each participant to find and walk along the way-points. The participants were put in a travel situation and instructed as follows (see *Figure 3*): *"You are at the central railway station in Vienna (Start 1) and want to visit the historic Belvedere Palace before traveling back home. To get there, first you have to buy a ticket at the ticket counter (waypoint 1 in Figure 3), then leave the luggage at the luggage lockers (waypoint 2). Afterwards go to the* *restroom (waypoint 3) and finally proceed to the stopping place of tram line D (waypoint 4)."*

To reduce the risk of cyber sickness the entire scenario was split into two parts. Each part was designed such that its completion was possible within 10 minutes under normal conditions. The first part starts at Start 1 and ends at waypoint 2 as shown in *Figure 3* by the red line, the second part starts at waypoint Start 2 and ends at waypoint 4 indicated by the orange line. An alternative route to the waypoint 4 using the escalators is also possible and can be seen in *Figure 3* by the orange dashed line. The green lines in *Figure 3* show intermediate goals where the agent is routed by direction arrows on the signs. If no sign could be seen the agent navigates from area to area to explore the infrastructure searching for signs or the target.

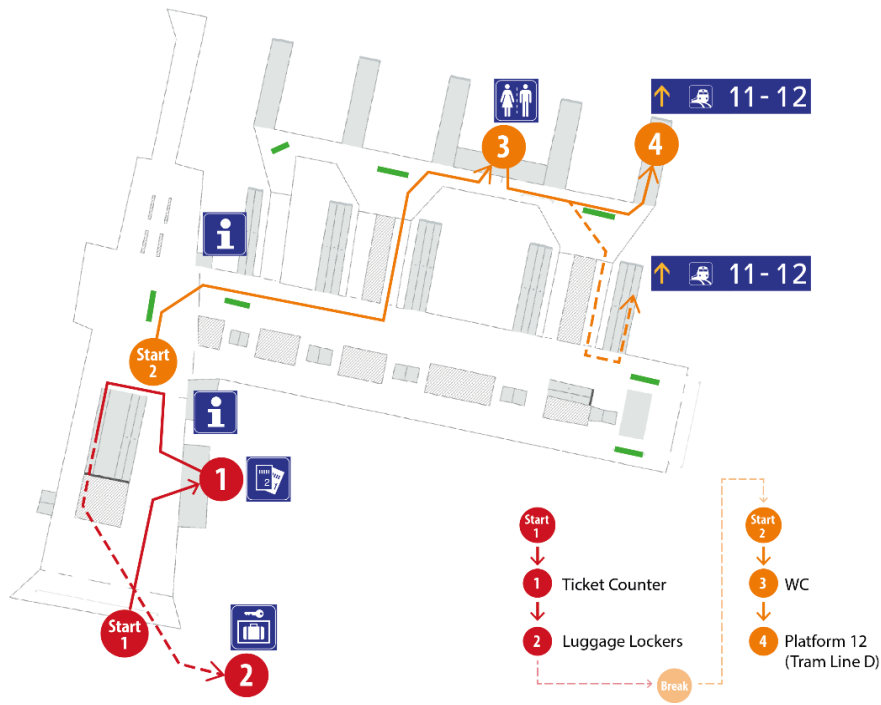

*Figure 3.* Waypoints and optimal route of the scenario for VR experiment and simulation

## 4.2   **Virtual Reality Experiments**

For the evaluation of our cognitive agent-based simulation, motion and visual attention data were collected in a controlled experiment using an immersive virtual reality environment in combination with a mobile eye tracking system for visual attention analysis (Schrom-Feiertag, Schinko, et al., 2014). After a short training session in the VR environment the participants were put in a travel situation and instructed with the scenario details. The experiment took place during the railway station's construction phase one year before its opening, therefore, no participant was familiar with the train station. For an enhanced immersion, virtual passers-by are simulated and an ambient soundscape was provided.

The validity of our VR environment for wayfinding research has been explored in (Bauer, Schneckenburger, et al., 2013) by conducting a case study with parallel test groups, exposing individuals to wayfinding exercises in the real world and the corresponding virtual world. The validation results showed that the perceived durations, egocentric distances and directions do not differ statistically significantly between the real and the virtual world.

From the experiments in the VR environment, we obtained accurate measurements on position, body orientation, viewing frustum and gaze of 20 participants (11 males and 9 females). These collected trajectories were used to evaluate the individuals' wayfinding behaviour and served for the validation of the simulation.

## 4.3　Simulation Results

*Figure 4* and *Figure 5* show the correlation between the obtained trajectories from the VR experiments (a) and the routes generated by the cognitive agent simulation (b). *Figure 4* shows the results of the first part of our experimental scenario, which includes the tasks of buying a ticket and finding the luggage lockers. While the ticket counter could be found in the experiment and in the simulation, finding the luggage lockers revealed much longer paths in the simulation where the entire hall was searched and some agents were not able to finish the task. The reason is, that the lockers are in a different storey of the building and only a small, inconspicuous sign points towards the stairs leading to the storey with the lockers. In the experiment, people quickly asked for help at the information counter which was not implemented in the simulation.

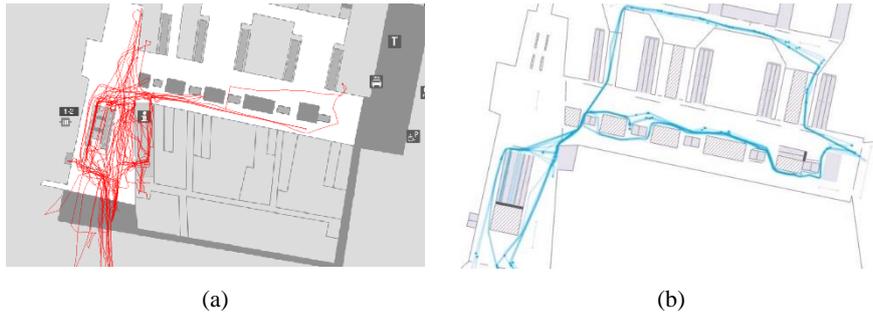

*Figure 4.* Trajectories of first part from VR experiment (a) and simulation (b).

*Figure 5* depicts the results of the second part of the scenario which includes the tasks of finding the WC and then the station of tram line D. The paths towards the WC, waypoint 3 in *Figure 3*, show a high correlation between the experiments and the simulation and the signage system was sufficient in both cases. Finding the tram line D deemed much more problematic as it requires either leaving the building and walking outside or walking along the whole platform 12 to find the tramway station outside the building. Given this task, the simulation stopped as it could not find the clues leading towards the station but also the test persons in the VR environment could not find the tram line without asking explicitly for it. As a fall back, we sent the simulated agent towards platform 12 at waypoint 4 which again could be easily found using the signage. Similar to the experiment, some agents used the stairs while some used the escalator to get to the floor with the platform.

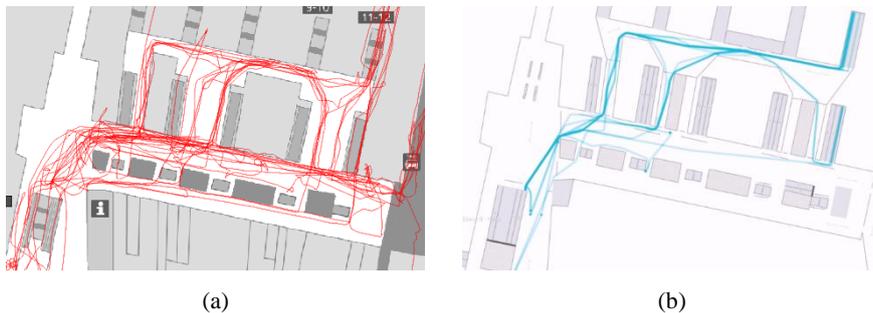

*Figure 5.* Trajectories of second part from VR experiment (a) and simulation (b).

The similarities in wayfinding between the experiment and the simulation seem promising, while the usage of info points and asking others for help needs further research. It might be, that this option was overused in the experiments as the test persons were given explicit instructions that they could ask for help in the VR experiments, which could be seen as an encouragement.

Furthermore, no other people queued at the info points which might be different in reality and decreases the probability for asking there for help.

## 5. CONCLUSIONS AND FUTURE RESEARCH

In this work, we proposed a novel modelling approach combining models for visual attention, human navigation and motion behaviour to build a cognitive software agent for the evaluation of indoor wayfinding systems. The agent-based pedestrian simulation has been applied to simulate realistic wayfinding behaviour given the infrastructure of the central railway station in Vienna. Empirical data collected from experiments using a VR environment were used for behavioural modelling and to validate simulation results of the same wayfinding scenario. The similarities of the trajectories between the experiment and the simulation seem promising and confirm the validity of the model. The experiments showed that the VR environment offers an innovative way to perform an evaluation of signage systems already at the planning stage. Especially in combination with eye-tracking it is also possible to obtain more fundamental knowledge about the wayfinding behaviour which in turn can be used to improve the models for attention and wayfinding behaviour.

At the present state of development, the simulation is ready for use, but the models are still based on simplifications. More fundamental questions can only be answered using a VR environment but the simulation can be seen as a useful complement to VR experiments. Since the implementation of such experiments is very time-consuming, the simulation offers an optimal solution to test a variety of scenarios which would not be possible experimentally with comparable effort.

The conducted experiments revealed alternative wayfinding strategies, like asking for the way if people are not sufficiently supported by the signage system. The next step will be the investigation and incorporation of such strategies in the navigation behavioural model.

Future research will also involve deeper analysis of eye tracking data to review the developed attention model on empirical data. In particular the dependency of the viewing frustum regarding the motion velocity needs to be investigated. Furthermore the viewing frustum of the agent is currently rendered straight ahead in the direction of movement. For a realistic perception of the surrounding environment additionally eye and head movements need to be considered.

## 6.    ACKNOWLEDGEMENTS

This work was part of the project MOVING and was funded by the research program line ways2go within the framework of the Austrian strategic initiative IV2Splus Intelligent Transport Systems and Services plus under the project number 835733.

## 7.    REFERENCES

Arthur, P. and R. Passini, 1990, *1-2-3 Evaluation and Design Guide to Wayfinding*, Technical Report, Public Works Canada.

Bauer, D., J. Schneckenburger, V. Settgast, A. Millonig, and G. Gartner, 2013, "Hands free steering in a virtual world for the evaluation of guidance systems in pedestrian infrastructures: Design and validation", Presented at the *Transportation Research Board 92nd Annual Meeting*.

Bierlaire, M., and T. Robin, 2010, "Pedestrians Choices", In H. J. P. Timmermans (Ed.), *Pedestrian behaviour: Models, data collection and applications*, Emerald Group Publishing, p. 1–26.

Daniel, K., A. Nash, S. Koenig, and A. Felner, 2010, "Theta*: Any-Angle Path Planning on Grids", *Journal Of Artificial Intelligence Research*, Volume 39, p. 533–579.

Foulsham, T., E. Walker, and A. Kingstone, 2011, "The where, what and when of gaze allocation in the lab and the natural environment", *Vision Research* 51, 17.

Gat, E., 1998, "On Three-Layer Architectures", in *Artificial Intelligence and Mobile Robots*, AAAI Press, Menlo Park.

Helbing, D., and P. Molnar, 1995, "Social Force Model for Pedestrian Dynamics", *Physical Review E*, 51, p. 4282–4286.

Henderson, J.M., and F. Ferreira, 2004, "Scene perception for psycholinguists", in Henderson, J.M., Ferreira, F., ed. *The Interface of Language, Vision, and Action: Eye Movements and the Visual World*, Psych. Press.

Hoogendoorn, S., P. Bovy, and W. Daamen, 2002, "Microscopic pedestrian wayfinding and dynamics modelling", in *Pedestrian and Evacuation Dynamics*, Springer, p. 124–154.

Hoogendoorn, S. P., and P. H. Bovy, 2004, "Pedestrian Route-Choice and Activity Scheduling Theory and Models", *Transportation Research, Part B: Methodological*, 38, p. 169–190.

Itti, L., C. Koch, and E. Niebur, 1998, "A model of saliency-based visual attention for rapid scene analysis", *Trans. Pattern Analysis & Machine Intelligence, 20, 11*.

Jovancevic-Misic, J., and M. Hayhoe, 2009, "Adaptive Gaze Control in Natural Environments", *The Journal of Neuroscience, 29, 19*.

Kielar, P. M., and A. Borrmann, 2016, "Modeling pedestrians' interest in locations: A concept to improve simulations of pedestrian destination choice", *Simulation Modelling Practice and Theory*, 61, p. 47–62.

Koh, W. L., and S. Zhou, 2011, "Modeling and Simulation of Pedestrian Behaviors in Crowded Places", *ACM Trans. Model. Comput. Simul.*, 21(3), p. 20:1–20:23.

Paletta, L., K. Santner, G. Fritz, H. Mayer, and J. Schrammel, 2013, "3D Attention: Measurement of Visual Saliency Using Eye Tracking Glasses", in *CHI '13 Extended Abstracts on Human Factors in Computing Systems*, CHI EA '13. New York, NY, USA, p. 199–204.

Raubal, M, 2001, "Human Wayfinding in Unfamiliar Buildings: A Simulation with a Cognizing Agent", *Cognitive Processing* 2 (3): 363–88.
Riche, N., M. Mancas, M. Duvinage, M. Mibulumukini, B. Gosselin, and T. Dutoit, 2013, "RARE2012: A multi-scale rarity-based saliency detection with its comparative statistical analysis", *Signal Processing: Image Communication*, *28*(6), p. 642–658.
Rothkopf, C., D. Ballard, and M. Hayhoe, 2007, "Task and scene context determine where you look", *Journal of Vision 7, 16.*
Schrammel, J., G. Regal, and M. Tscheligi, 2014, "Attention Approximation of Mobile Users Towards Their Environment", in *CHI '14 Extended Abstracts on Human Factors in Computing Systems*, New York, NY, USA, p. 1723–1728.
Schrammel, J., E. Mattheiss, S. Döbelt, L. Paletta, A. Almer, and M. Tscheligi, 2011, "Attentional Behavior of Users on the Move Towards Pervasive Advertising Media", in *Pervasive Advertising*.
Schrom-Feiertag, H., C. Schinko, V. Settgast, and S. Seer, 2014, "Evaluation of Guidance Systems in Public Infrastructures Using Eye Tracking in an Immersive Virtual Environment", in *Proceedings of the 2nd International Workshop on Eye Tracking for Spatial Research*, Vienna, p. 62–66.
Sun, R., 2008, "Introduction to Computational Cognitive Modeling", *Cambridge Handbook of Computational Psychology*, 3–19.
Timmermans, H. J. P., 2009, *Pedestrian Behavior: Models, Data Collection and Applications*, Emerald Group Publishing, Bingley, UK.
Wiener, J. M., S. J. Büchner, and C. Hölscher, 2009, "Taxonomy of Human Wayfinding Tasks: A Knowledge-Based Approach", *Spatial Cognition & Computation*, 9(2), p. 152–165.